# Refractometry of birefringent materials at Brewster angle

E.A.Tikhonov, A.K. Lyamets

Institute of Physics, Nat.Acad.Sci. of Ukraine, 03028, Kyiv, Science Avenue 46, e-mail:etikh@live.ru

**Abstract** - A refractometry taking measuring of Brewster angles for birefringent materials is proposed. The technique is based on the angular scanning of the reflected power of p-polarized light and the determination of the two specific minimums at the Brewster angles for the two orthogonal orientations of the optical axis of the material. In these configurations, the testing light does not experience the birefringent in the material, assuming the state of ordinary or extraordinary rays alternately to orientation of the optical axis. The origin and influence on the measurement of the nonzero residual power at Brewster angle are accentuated.

## 1. INTRODUCTION

It is well known that the minimal reflected power at the Brewster angle is specified by the transverse structure of the oscillations of the electromagnetic field. In our works [1,2,3] physical and metrological aspects of the laser measuring the refractive index (RI) at Brewster angle for the isotropic materials were presented. Such factors as absorption of the testing light by the material, internal and surface light scattering on the tested and adjacent surfaces were taken into account. The advantages of Brewster refractometry is in its suitability to RI measurement any materials with arbitrary shape, but just in the presence of a single surface of optical quality. With this surface, a single measurement should be done to determine the absolute value of Brewster angle. Given paper presents the application of the above-mentioned technique for birefringent materials.

## 2. CORE OF BREWSTER REFRACTOMETRY

Measurement of the absolute Brewster angle value in the vicinity of the fuzzy minimum of the reflected power of p-polarized light (TH-wave) was performed with computer control of the turning angles of this surface with a timed reading of the reflected power by step of 1 arc. minute [1,2]. The difference in the measurement of the RI of uniaxial birefringent materials at Brewster angle compared to the procedure for isotropic materials is reposed by the need to combine the optical axis of the material with its test plane. After this combining, it is asked to adjust parallel the turning axis and optical axis of the birefringent material. Now the condition for determining the Brewster angle for an ordinary beam, formed by a TH-wave, becomes feasible.

In our measurements the electric field vector of the TH-wave remained always in the horizontal plane (specific set-up). Excitation of extraordinary wave and determining of the inherent Brewster angle requires the parallel adjustment this vector and the optical axis of the material. However, converting the optical axis to the horizontal plane and subsequent measurements make only coplanar configuration with a variable angle of its intersection ( x) in the all vicinity of the Brewster angle scanning. The plane defined by these crossed vectors is orthogonal to the axis of the angular scan. For the same time, only the projection electric field vector of the TH-wave proportional to cos (x) stays parallel to the optical axis, while the orthogonal projection is proportional to sin (x). In the dipole approximation, excitation of the polarization wave from the sin(x)-component is absent because of the transverse field structure. So this adjustment implies the partial fulfillment of the requirements for the excitation only extraordinary wave and the correct determination of Brewster angle will be checked out experimrntally.

As objects of testing and certification of the method about the accuracy, we used transparent ScotchTape polymer films with adhesion to substrates, $\lambda/4$-wave quartz plates and muscovite mica samples with an atomic-clean surface. The Scotch Tape film contains a basic polymer layer and an adhesive-sensitive layer that adheres on the substrate surface bythe pressure-sensitive adhesive layer.

It is known that deformations of isotropic amorphous materials are accompanied by the appearance of optical anisotropy from Brewster time as photoelasticity effect and now [4]. The used Scotch Tape adhesive film had a thickness 50µm and wideness 18mm. The film is characterized by stable optical anisotropy resulted from the technology of the similar film production in the state of plastic deformation [5,6]. The directional deformation with stretching of an isotropic film leads to the formation of optical anisotropy due to the partial alignment of the polymer molecules along the direction of stretching.

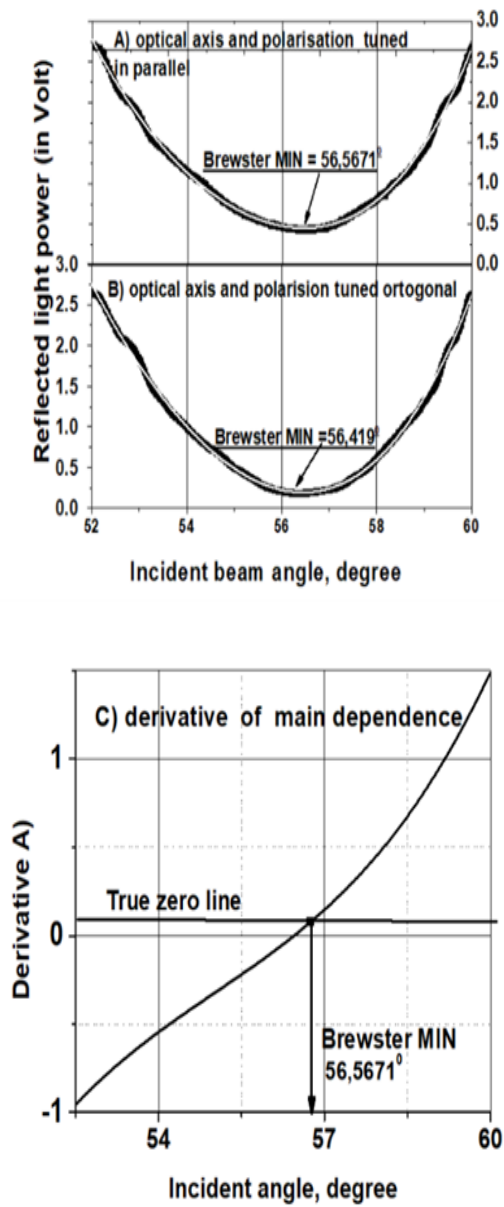

film around an orthogonal axis. These experiments discovered that the direction of the optical axis oriented along the longitudinal direction (the direction of twist of the film roll). This allowed identifying the direction of plastic deformation with the partial alignment of polymer molecules [5]. The final conclusion about the direction of the optical axis and the value of birefringence for the tested polymer film will be performed at quantitative measurements of RI by the proposed method. The results of measuring the angular distribution of the reflected power of a He-Ne laser at a wavelength of 632.8 nm with a given justification of the technique are shown in Fig.1a,b,c. The initial data for these graphs are registered in tabular format. The format allows determining the angular position of the minimum of reflected power with sub-minute accuracy and, accordingly, the numerical values of the refractive indices for ordinary and extraordinary waves under support of the basic Fresnel theory formula $\tan(\varphi_{Brw}) = n_{e,o}$ under conditions of a uniform angular distribution of the noise power [1].

However, for better clarity of the results, we present their graphical manifestation with support of the "OriginPro" program. In Fig.1a, b the angular dependences of the reflected HeNe laser power in the vicinity Brewster angle for the two orthogonal orientations optical axis (ScotchTape film) is shown. The image of reference points by a 9-pixel symbol at a step of 1 arc-minute on selected interval these dependencies forms a wide continuous strip. Linear fourth power polynomial regression (white line) of the obtained results was inserted along with standard deviation not worse than 0.99. However, the residual power of the signal at the Brewster angle is not zero! Therefore, the derivative with respect to the scan angle (Fig. 1c) at the intersection point with the original abscissa does not indicate the true value of the Brewster angle. But because there is a zero minimum for the derivative, the constant component of the residual noise disappears. Then "non-zero" Brewster minimum from the table format and the zero minimum for the derivative curve have the same numerical values. (Fig.1a,b, c).

The appearance of residual power at the Brewster angle contradict to the Fresnel theory. But in fact, there are a number of physical reasons for the existence of nonzero residual power: the optical roughness of the reference surface (light scattering), the final degree of linear polarization of the testing beam and the final accuracy of its alignment with the plane of incidence, as well as the

Fig.1a,b,c Angular dependences of TH-wave reflection power for parallel and orthogonal orientations of the optical axis for ScotchTape film (a, b) and derivative respect to scan angle for the dependence A) atan(j) indication of correct determination of the Brewster angle and exclusion of residual power.

This state of deformation with the resulting molecular ordering resembles the state of homogeneous orientation of nematic liquid crystals and can be described using the order parameter [4].

## 2. DETERMINATION IR AT BIREFRINGENCE

Initial testing the birefringence and the angular orientation of the optical axis in the Scotch Tape plane in the crossed polarized arrangement was shown the behavior similar to that of $\lambda/4$-quartz plate. The four states of light transmission "open/closed" is registered for a full $2\pi$-turn of the contribution of the test beam, reflected diffusely and coherently to the back surface (in the case of films, plates [2, 3]. When comparing the actual levels of the residual power of a given film (without optical treatment) with the atomic surface of mica (should be further) the difference in residual power will be several orders of magnitude. The reason of the concern for residual power is a decrease in the accuracy of true Brewster angle measurements the due to a possible displacement of the angular position of the minimum because the contribution of noise power varying in angle [1,2].

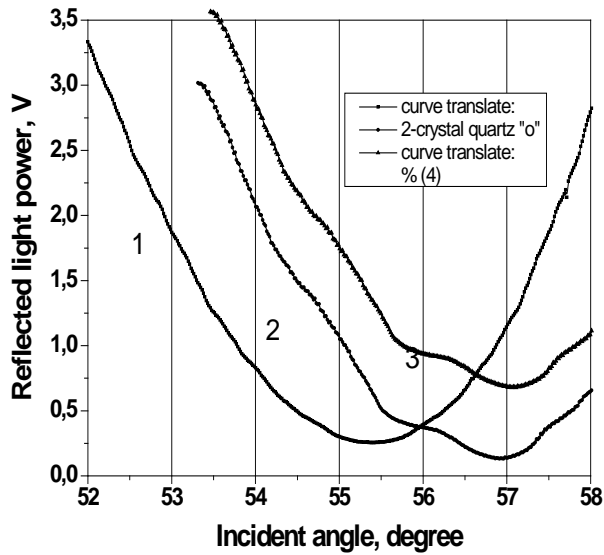

Fig.2. Determination of the Brewster angles and RI for fused quartz (1) and RI = $n_{e,o}$ crystalline quartz (2.3) from angular dependences of the reflected power of the TH-wave 632.8 nm, thickness $\lambda/4$ - quartz is plate 2.17mm (curves2.3 translated vertically for best seen). of the residual noise disappears.

Testing of the new refractometry about reliability was carried out by comparing the obtained values of RI of ordinary/extraordinary rays with known data for such RI of crystalline quartz. The level of residual power at using a quartz plate with a pair of surfaces of optical quality might limit the accuracy of measurement, so reflections from the back surface should be removed (by immersion, using a wedged or an absorbing layer). The measured values of the birefringence on this film and quartz plate turned out comparable 0.005÷0,010 at various magnitudes of RI. The absolute values of RI for quartz from the broadly known data are equal for $n_{e,o}$ = 1.544 and 1.536, respectively, and differ in the 3rd sign from the values on the yellow sodium line (the effect of dispersion). Measured graphic dependencies with numerical values of $n_{e,o}$ for $\lambda$= 632,8nm are shown in Fig.2. For comparison, a typical dependence for fused silica is given with a Brewster angle, which determines the value RI = 1.4505. Under the same conditions about the power level of the testing laser beam, the residual power levels for quartz plates (2.3) and fused silica were comparable.

The influence of the optical quality of the surface on the level of residual power at Brewster angle can be proved by comparing the results shown in Fig. 1.2 with similar ones for mica plate such as muscovite (Fig.3). Muscovite belongs to the monoclinic system and is characterized by the presence of 2 optical axes [7]. Both axes lie in a plane orthogonal to the cleavage plane. The latter is ideal for the smallness of surface defects that create diffuse light scattering and, accordingly, the residual power at Brewster angle.

The classification of emerging beams for 2-axis crystals is unknown. Therefore, the used optical configuration for finding the minimum of the reflected power did not correspond to any of the known single-beam excitation configurations. With a small ratio of the thickness of the plate/diameter of the laser beam T/Ø≈0.08, the emerging beams mutually overlap, so that with angular scanning of the reflected power, the first feature manifested itself in the form of suppression of 2-beam interference at an angle of $32^0$. This effect can be explained by the occurrence of orthogonal polarization between two interfering beams due to an accidental phase change (reflection from the back side in this experiment was not suppressed) with a synchronous change in the modulation depth of the 2 beam interference. The original Brewster minimum was detected at an angle of $\varphi_{Brw}$ = $58.166^0$ with a value of IR = 1.610. Earlier certain values of IR mica along the optical axes are the following: $n_g$ = 1.613 ÷ 1.596, $n_m$ = 1.607 ÷ 1.596, $n_p$ = 1.569 ÷ 1.561 [7], so the obtained value of RI falls within the range of values $n_{brw} \approx n_m$. But presented on fig.3. dependence for the muscovite mica is not a singular one because changes at the turn of the tested plane around an orthogonal axis (details of this research will be published in the relevant paper).

The residual power at the Brewster angle for muscovite determined by the noise level of the recording system and outside scattering light ≈10 mV, and was lower compared to the power of the reflected laser beam power outside Brewster angle more than 3 orders of magnitude (Fig. 3). Therefore, it is not necessary to associate the nature of the residual power with reflection from a boundary layer of matter (Drude's model [2]).

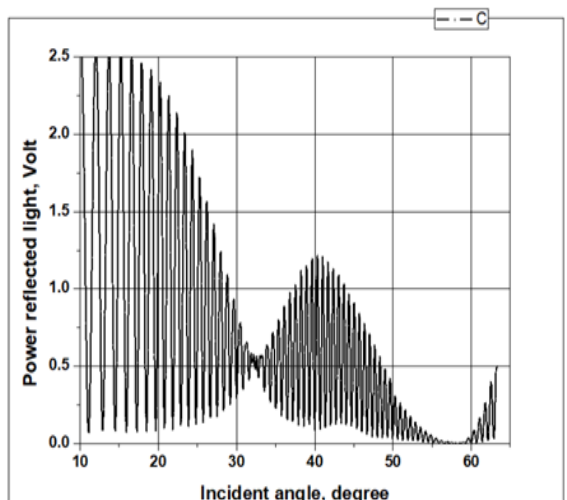

Fig.3. Determination of the Brewster angle for muscovite mica along the cleavage plane at a TH-wavelength of 632.8 nm, the interference pattern is due plane-parallel plate at thickness is 80 microns.

**4. FINDING**

In conclusion, we note that this report presents a new method for determining refractive indices for uniaxial birefringent materials using Brewster refractometry method. The principal difference this technique in compare to isotropic materils is the specific requirement of preparing the reference plane of the birefringent material for measurements: this plane should be combined with the optical axis of the material and specifically oriented relative to axis of the angular scanning. The other factors that touch an accuracy of measurement remain the same as for measurements of isotropic materials [2].